\documentstyle[11pt,epsf]{article}

\def\beq{\begin{eqnarray}}
\def\eeq{\end{eqnarray}}
\def\bea{\begin{eqnarray*}}
\def\eea{\end{eqnarray*}}

\def\NPB#1#2#3{Nucl. Phys. {\bf B#1} (#2) #3}
\def\PLB#1#2#3{Phys. Lett. B {\bf #1} (#2) #3}

\def\PRD#1#2#3{Phys. Rev. {\bf D#1} (#2) #3}

\def\PRL#1#2#3{Phys. Rev. Lett. {\bf #1} (#2) #3}
\def\PREP#1#2#3{Phys. Rep. {\bf #1} (#2) #3}

\def\centeron#1#2{{\setbox0=\hbox{#1}\setbox1=\hbox{#2}\ifdim
\wd1>\wd0\kern.5\wd1\kern-.5\wd0\fi
\copy0\kern-.5\wd0\kern-.5\wd1\copy1\ifdim\wd0>\wd1
\kern.5\wd0\kern-.5\wd1\fi}}
\def\ltap{\;\centeron{\raise.35ex\hbox{$<$}}{\lower.65ex\hbox{$\sim$}}\;}
\def\gtap{\;\centeron{\raise.35ex\hbox{$>$}}{\lower.65ex\hbox{$\sim$}}\;}
\def\gsim{\mathrel{\gtap}}
\def\lsim{\mathrel{\ltap}}

\def\slashchar#1{\setbox0=\hbox{$#1$}           % set a box for #1
   \dimen0=\wd0                                 % and get its size
   \setbox1=\hbox{/} \dimen1=\wd1               % get size of /
   \ifdim\dimen0>\dimen1                        % #1 is bigger
      \rlap{\hbox to \dimen0{\hfil/\hfil}}      % so center / in box
      #1                                        % and print #1
   \else                                        % / is bigger
      \rlap{\hbox to \dimen1{\hfil$#1$\hfil}}   % so center #1
      /                                         % and print /
   \fi}                                        %

\def\singleandthirdspaced{\baselineskip=\normalbaselineskip\multiply
    \baselineskip by 130\divide\baselineskip by 100}
\def\singlespaced{\baselineskip=\normalbaselineskip}

\newcommand{\newc}{\newcommand}
\newc{ \Ni         }{ {\tilde N}_i }
\newc{ \Nj         }{ {\tilde N}_j }
\newc{ \NI         }{ {\tilde N}_1 }
\newc{ \NII        }{ {\tilde N}_2 }
\newc{ \NIII       }{ {\tilde N}_3 }
\newc{ \NIIII      }{ {\tilde N}_4 }
\newc{ \Ci         }{ {\tilde C}_i }
\newc{ \Cj         }{ {\tilde C}_j }
\newc{ \CI         }{ {\tilde C}_1 }
\newc{ \CII        }{ {\tilde C}_2 }
\newc{ \CIp        }{ {\tilde C}^{+}_1 }
\newc{ \CIm        }{ {\tilde C}^{-}_1 }
\newc{ \Cip        }{ {\tilde C}^{+}_i }
\newc{ \Cim        }{ {\tilde C}^{-}_i }
\newc{ \Cjp        }{ {\tilde C}^{+}_j }
\newc{ \Cjm        }{ {\tilde C}^{-}_j }
\newc{ \G          }{ {\tilde G} }
\newc{ \XI         }{ {\tilde X}_1 }
\newc{ \XII        }{ {\tilde X}_2 }
\newc{ \eL         }{ {\tilde e}_L }
\newc{ \eR         }{ {\tilde e}_R }
\newc{ \veL        }{ {\tilde \nu} }
\newc{ \SG         }{ {\tilde \gamma} }
\newc{ \SZ         }{ {\tilde Z} }
\newc{ \gmu        }{ \gamma^{\mu} }
\newc{ \gnu        }{ \gamma^{\nu} }
\newc{ \gamone      }{ \gamma_{1} }
\newc{ \gamtwo      }{ \gamma_{2} }
\newc{ \gamboth      }{ \gamma_{1,2} }
\newc{ \dL         }{ \tilde d_L }
\newc{ \dR         }{ \tilde d_R }
\newc{ \uL         }{ \tilde u_L }
\newc{ \uR         }{ \tilde u_R }
\newc{ \slepton    }{ \widetilde \ell }
\newc{ \M          }{ {\cal M} }
\newc{ \ra         }{ \rightarrow }
\newc{ \ltilde     }{ {\tilde \ell} }
\newc{ \nutilde    }{ {\tilde \nu} }
\newc{ \lLstar     }{ { \tilde \ell}_L^* }
\newc{ \lRstar     }{ { \tilde \ell}_R^* }
\newc{ \snu        }{ { \tilde \nu} }
\newc{ \snustar    }{ { \tilde \nu}^* }
\newc{ \nubar      }{ \overline{ \nu } }
\newc{ \muL        }{ { \tilde \mu}_L }
\newc{ \muR        }{ { \tilde \mu}_R }
\newc{ \tauL       }{ { \tilde \tau}_L }
\newc{ \tauR       }{ { \tilde \tau}_R }
\newc{ \h          }{ { h^0 } }
\newc{ \Et         }{ { \slashchar{E}_T } }
\newc{ \Etot         }{ { \slashchar{E} } }
\newc{ \spt         }{ { \slashchar{p}_T } }
\newc{ \Etcut      }{ { \slashchar{E}_T^{\rm cut} } }
\newc{ \sigbreff   }{ \sigma \times {\rm BR} \times {\rm EFF} }
\newc{ \eegg       }{ {ee\gamma\gamma} }
\newc{ \GeV        }{ {\rm GeV} }
\newcommand{\sss}{\scriptscriptstyle}

\newcommand{\sel}{\tilde{e}_{\sss L}}
\newcommand{\ser}{\tilde{e}_{\sss R}}
\newcommand{\sdr}{\tilde{d}_{\sss R}}
\newcommand{\sur}{\tilde{u}_{\sss R}}
\newcommand{\sSr}{\tilde{s}_{\sss R}}
\newcommand{\sCr}{\tilde{c}_{\sss R}}
\newcommand{\sdl}{\tilde{d}_{\sss L}}
\newcommand{\sul}{\tilde{u}_{\sss L}}

\newcommand{\gluino}{\tilde{g}}
\newcommand{\squark}{\tilde{q}}
\newcommand{\mplus}{m_+}
\newcommand{\mminus}{m_-}
\newcommand{\mf}{m_f}
\newcommand{\Phibar}{{\overline \Phi}}
\newcommand{\meps}{m_\epsilon}

\begin{document}

\begin{titlepage}
\begin{flushright}
{\large
hep-ph/9608224}
\end{flushright}

\vskip 1.2cm

\begin{center}
{\LARGE\bf Generalized messengers of supersymmetry breaking}

{\LARGE\bf and the sparticle mass spectrum}

\vskip 2cm

{\large
 Stephen P. Martin\footnote{{\tt spmartin@umich.edu}}
} \\
\vskip 4pt
%Hail! to the victors valiant, Hail!...
{\it Randall Physics Laboratory\\
     University of Michigan\\
     Ann Arbor MI 48109--1120 } \\

\vskip 1.5cm

\begin{abstract}

We investigate the sparticle spectrum in models of gauge-mediated
supersymmetry breaking. In these models, supersymmetry is
spontaneously broken at an energy scale only a few orders of
magnitude above the electroweak scale. The breakdown of supersymmetry
is communicated to the standard model particles and their
superpartners by ``messenger" fields through their ordinary gauge
interactions. We study the effects of a messenger sector in which
the supersymmetry-violating $F$-term contributions to messenger
scalar masses are comparable to the supersymmetry-preserving ones. We
also argue that it is not particularly natural to restrict attention
to models in which the messenger fields lie in complete $SU(5)$ GUT
multiplets, and we identify a much larger class of viable models.
Remarkably, however, we find that the superpartner mass parameters in
these models are still subject to many significant contraints.

\end{abstract}

\end{center}

\vskip 1.0 cm

\end{titlepage}
\setcounter{footnote}{0}
\setcounter{page}{2}
\setcounter{section}{0}
\setcounter{subsection}{0}
\setcounter{subsubsection}{0}

%%%%%%%%%%%%%%%%%%%%%%%%%%%%%%%%%%%%%%%%%%%%%%%%%%%%%%%%%%%%%%%%%%%%%%%
\singleandthirdspaced
\section*{1. Introduction}
\indent

The masses of the superpartners of the Standard Model (SM) particles
should not greatly exceed the TeV scale if supersymmetry is to solve
the hierarchy problem associated with the ratio $M_Z/M_{\rm Planck}$.
However, this fact by itself
tells us surprisingly little about the scale $\Lambda_{\rm SUSY}$
at which supersymmetry is ultimately broken. It is also necessary
to have an understanding of the mechanism by which supersymmetry
breaking is communicated from its original source to the fields
of the Minimal Supersymmetric Standard Model (MSSM). If gravitational
or other Planck-suppressed
interactions communicate supersymmetry breaking, then $\Lambda_{\rm SUSY}$
is perhaps $10^{11}$ GeV or so. While this scenario has received the
most attention in the last decade, it is hardly inevitable. Another
possibility \cite{oldmodels,GaugeMediated}
is that the ordinary gauge interactions are responsible
for communicating supersymmetry breaking to the MSSM through their couplings to
a messenger sector of chiral superfields,
which in turn couple
directly or indirectly to the fields which break supersymmetry.

In the ``minimal" model of gauge mediated supersymmetry breaking (GMSB)
\cite{GaugeMediated}, all of the
soft supersymmetry-breaking interactions of the MSSM
are determined by just a few free parameters.
Perhaps the most attractive feature of this type of model
is that the masses generated for squarks and sleptons with
the same $SU(3)_C\times SU(2)_L \times U(1)_Y$
quantum numbers are automatically degenerate, so that
flavor-changing neutral currents are suppressed without additional
assumptions. This feature depends only on the fact that ordinary
gauge interactions are flavor-blind, and will be true in a much
larger class of models than just the minimal GMSB model.

This class of models has another feature which may allow it to be dramatically
confirmed at existing or currently planned collider facilities. Because local
supersymmetry is spontaneously broken at a
relatively low scale, the lightest supersymmetric particle is
the gravitino (the spin $3/2$ superpartner of the graviton),
with a mass that is entirely irrelevant for collider kinematics
(but not for cosmology\cite{cosmoconstraints}).
The next-to-lightest supersymmetric particle (NLSP) can therefore
decay into its SM partner and the gravitino.
In the case that the lightest neutralino ($\NI$) is
the NLSP, one has the interesting decay\cite{Fayet,decay}
$\NI\rightarrow\gamma\G$
as long as the photino content of $\NI$ is non-zero.
The decay length for this process depends on the ultimate scale of
supersymmetry breaking $\Lambda_{\rm SUSY}$, according to
\beq
\Gamma(\NI \rightarrow \gamma\G) =
{\kappa_{1 \gamma} \over 16 \pi}
{m_{\NI}^5 \over \Lambda_{\rm SUSY}^4}
\eeq
where $\kappa_{1\gamma} = |N_{11} \cos\theta_W + N_{12}\sin\theta_W |^2$
(in the notation of \cite{HaberKane}) is the photino content of $\NI$.
Since in GMSB the typical
$F$-term responsible for supersymmetry breaking can correspond to
$\Lambda_{\rm SUSY}$ of order $10^2$
or $10^3$ TeV, it is quite possible that this decay can occur
(at least a significant fraction of the time)
inside a typical detector,
with many interesting phenomenological
consequences\cite{Fayet,decay,SWY,DDRT,AKKMM1,KNW,DTW,AKKMM2,BKW,LN}.
If it is sufficiently heavy $\NI$ can also have decays into $Z\G$ and $h\G$,
with decay widths which suffer, however, from very
strong kinematic suppression\cite{AKKMM2}.

Recently it was pointed out\cite{DDRT,AKKMM1}
that a single $ee\gamma\gamma+\Et$
event \cite{event} observed at CDF could be naturally
explained\footnote{The event can also be explained in the usual
MSSM framework without a light gravitino, if parameters
are chosen so that the radiative 1-loop
decay $\NII\rightarrow \NI \gamma$ dominates\cite{AKKMM1,AKKMM3}.
The parameter space in which this can occur will be largely but not
entirely explored at LEP161 and LEP190.}
by GMSB (and other theories with a light gravitino).
This event had an energetic electron and positron, two energetic
photons each with pseudorapidity $|\eta |<1$ and transverse energy
$E_T > 30$ GeV, and large missing transverse energy
$\Et > 50$ GeV. The SM and detector backgrounds for such events
are reputed to be extremely small. This event can
be explained by GMSB
as either selectron pair production or
chargino pair production, but only
if $\NI$ is the NLSP, and if $\Lambda_{\rm SUSY}$ is less than about
$10^3$ TeV. These are not automatic consequences of all models, and therefore
will give
(if taken seriously, which clearly should not be considered mandatory!)
non-trivial theoretical constraints.

Moreover, the discovery
signatures of supersymmetry with a prompt decay $\NI\rightarrow\gamma\G$
are so spectacular that it is possible to set quite strong bounds
even with existing Tevatron data. In contrast to the usual supersymmetry search
strategies, one can obtain a very high detection efficiency
at the Tevatron for the inclusive signal $\gamma\gamma+X+\Et$
with suitable cuts on the transverse energy and isolation of the
photons, and on the total missing transverse energy.
In \cite{AKKMM2} it was argued that with
the present 100 pb$^{-1}$ of data at the Tevatron, it should be
possible to exclude a lightest chargino ($\CI$) mass
up to 125 GeV and neutralino
masses up to about 70 GeV, assuming gaugino
mass ``unification" relations as in the minimal GMSB model. In this paper
we will discuss other models which do not share this feature.
Even when all assumptions about gaugino mass relations are abandoned,
however, it was argued in \cite{AKKMM2} that one can still find a model
independent bound $m_{\CI} > 100 $ GeV as long as $m_{\NI}>50$ GeV
(to supply energetic photons) by exploiting the inclusive
$\gamma\gamma+X+\Et$ signal. These bounds are quite competitive with
and somewhat complementary to what
can be done at LEP upgrades.
However, it should be kept in mind that
these bounds all assume that the decay $\NI\rightarrow\gamma\G$ occurs
within the detector 100\% of the time. This is not necessary, even to
explain the CDF $ee\gamma\gamma+\Et$ event, which only requires that some
non-negligible fraction of $\NI$ decays occur within the detector.
If most decays occur outside the detector, then one would expect
many more single photon events than diphoton events, with
unfortunately a much larger SM background, and much more difficult
challenges for simulation studies. Thus for example the
discovery mode at LEP2 from $e^+ e^-\rightarrow \NI\NI$
could be predominantly $\gamma\Etot$ rather than $\gamma\gamma\Etot$.
We should also note that in a significant fraction
of the models to be studied in this paper, $\NI$ cannot be the NLSP
anyway unless it is higgsino-like.

While the minimal model of GMSB is quite elegant and can explain
the CDF $ee\gamma\gamma+\Et$ event, it is important to consider
what all the related alternatives might be, especially in setting
discovery and exclusion strategies.
Future phenomenological studies
should therefore take into account the full richness of model-building
possibilities, which undoubtedly extend far beyond the minimal
GMSB model and in several different
directions\cite{DGP1,DGP2,Faraggi,CKN,DNS,DDR}.
In this paper we will begin to explore a few such possibilities.
In section 2 we develop the
formalism for arbitrary messenger sector field content including
the effects of arbitrary masses (from scalar VEVs and
$F$-term breaking) in the
messenger sector. In section 3
we will examine the discrete model space allowed by
generalizing the particle content of the messenger sector to
include possibilities which do not form complete GUT multiplets.
We will argue that it is not particularly unnatural or even
inelegant to consider such generalizations.
These effects serve to considerably enlarge the
available parameter space, but in section 4 we show that some strong
model-independent statements can still be made, and the GMSB models
retain a distinct character even without taking into
account the possibility of discovery modes involving decays
into the gravitino.

\section*{2. Beyond the minimal model}
\indent

In this section we consider a slightly generalized treatment of
the minimal model of GMSB.
The messenger sector consists of a set of chiral superfields
$\Phi_i,\Phibar_i$ which transform as a vector-like representation of the
MSSM gauge group. The supersymmetry breaking mechanism is parameterized
by a (perhaps not fundamental) chiral superfield $S$, whose
auxiliary component $F$ is assumed to acquire a VEV. The messenger
fields couple to $S$ according to the superpotential
\beq
W= \lambda_i S \Phi_i\Phibar_i
\eeq
(Here we have assumed that the messengers
obtain their masses only from coupling to a single chiral
superfield $S$; we will comment briefly on the effects of
relaxing this assumption below. With this assumption, a possible
coupling matrix $\lambda_{ij}\Phi_i\Phibar_j$ can always be diagonalized
as shown.)
In the minimal model of GMSB \cite{GaugeMediated},
$\Phi_i$ and $\Phibar_i$ consist of chiral superfields transforming as a
${\bf 5 } +{\bf\overline 5}$ of $SU(5) \supset SU(3)_C\times SU(2)_L
\times U(1)_Y$. This choice is sufficient to give masses to all of the
MSSM scalars and gauginos.

In the following we will use the same symbol for $S$ and $F$ and for
their VEVs.
The fermionic components of $\Phi_i$ and $\Phibar_i$ obtain a Dirac
mass equal to $\lambda_i S$. Their scalar partners have a (mass)$^2$
matrix equal to
\beq
\pmatrix{|\lambda_i S|^2 & \lambda_i F \cr
         \lambda_i^* F^* & |\lambda_i S|^2 \cr}
\label{scalmatrix}
\eeq
with eigenvalues $ |\lambda_i S|^2 \pm |\lambda_i F|$.
The supersymmetry violation apparent in this spectrum
is then communicated to the MSSM sector
via the ordinary gauge interactions of $\Phi_i$ and $\Phibar_i$.

\begin{figure}[t]
\centering
\epsfxsize=3.1in
\hspace*{0in}
\epsffile{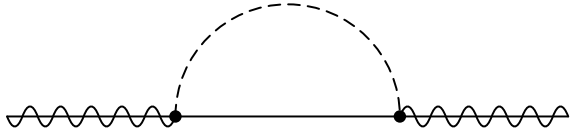}
\caption{Contribution to MSSM gaugino masses from messenger field loops}
\label{oneloopfigure}
\end{figure}

The gauginos of the MSSM obtain their masses at one loop from
the diagram shown in Fig.~\ref{oneloopfigure}.
%%%%%%%%%%%%%%%%%%%%%%%%%%%%%%%%%%%%%%%%%%%%%%%%%%%%%%%%%%%%%%%%%%%%%
%%%%%%%%%%%%%%%%%%%%%%%%%%%%%%%%%%%%%%%%%%%%%
The particles in the loop are the messenger fields.
Evaluating this graph one finds that the MSSM gaugino mass parameters
induced are:
\beq
M_a = {\alpha_a\over 4\pi} {F\over S}
\sum_i n_a(i) g(x_i) \qquad\qquad (a=1,2,3)
\label{gauginomasses}
\eeq
where
\beq
x_i = |F/\lambda_i S^2|
\eeq
for each messenger coupling $\lambda_i$ and
\beq
g(x) =
{1\over x^2}[{(1+x)\log (1+x) } +
{(1-x)\log (1-x) }]\> .
\eeq
In eq.~(\ref{gauginomasses}),
$n_a(i)$ is the Dynkin index for the
pair $\Phi_i,\Phibar_i$ in a normalization where $n_a=1$ for
${\bf N} + {\bf \overline N}$ of $SU(N)$. We always use a GUT
normalization for $\alpha_1$ so that $n_1 = {6\over 5} Y^2$ for
each messenger pair with weak hypercharge $Y=Q_{\rm EM} - T_3$.
The variable $x_i$ must lie in the range $0<x_i<1$, with
the upper limit coming from the requirement that the lighter
scalar messenger has positive (mass)$^2$. The minimal
${\bf 5}+{\bf\overline 5}$ model has $\sum_i n_1(i) =
\sum_i n_2(i) = \sum_i n_3(i) = 1$.
Since the function
$g(x)$ obeys $g(0)=1$, in the small $x_i$ limit one recovers
the result $M_a = (\alpha_a/4\pi)F/S$
of \cite{GaugeMediated} for the minimal model.

\begin{figure}[t]
\centering
\epsfxsize=3.85in
\hspace*{0in}
\epsffile{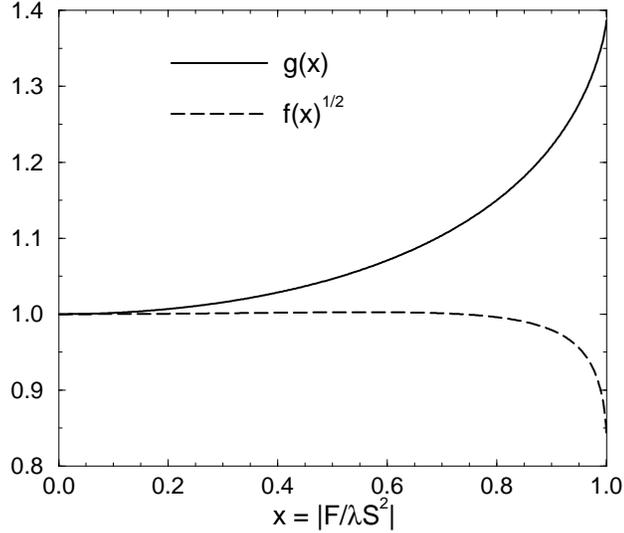}
\caption{The functions $g(x)$ and $\protect\sqrt{f(x)}$
%%   Geez LaTeX sucks eggs!!!!!    ^^^^^^^^
%% What the hell kind of buggy crap is THIS!?!?!?!?!?
described in the text.}
\label{fgplot}
\end{figure}

For larger $x$ the expansion
\beq
g(x) = 1 + {x^2\over 6} + {x^4 \over 15} + {x^6 \over 28} +\ldots
\eeq
gives good accuracy except near $x=1$.
The function $g(x)$ is graphed in Figure \ref{fgplot},
and can be seen to
increase monotonically
%%%%%%%%%%%%%%%%%%%%%%%%%%%%%%%%%%%%%%%%%%%%%%%%%%%%%%%%%%%%%%%%%%%%%
%%%%%%%%%%%%%%%%%%%%%%%%%%%%%%%%%%%%%%%%%%%%%
with $x$, reaching a maximum value $g(1)=2\log 2\approx 1.386$.
It is sometimes convenient to write
$M_a = {\alpha_a \over 4 \pi} \Lambda_{Ga}$ where
\beq
\Lambda_{Ga} = {F\over S} \sum_i n_a(i) g(x_i)
\qquad\qquad a=1,2,3
\eeq
parameterizes the possible effects of a non-minimal
messenger sector and non-negligible $x_i$. In general
one finds
\beq
{F\over S} N_a \leq \Lambda_{Ga} \leq 1.386 {F\over S} N_a
\eeq depending on $x_i$, where
\beq
N_a = \sum_i n_a(i) .
\eeq

\begin{figure}%[t]
\centering
\epsfxsize=5.8in
\hspace*{0in}
\epsffile{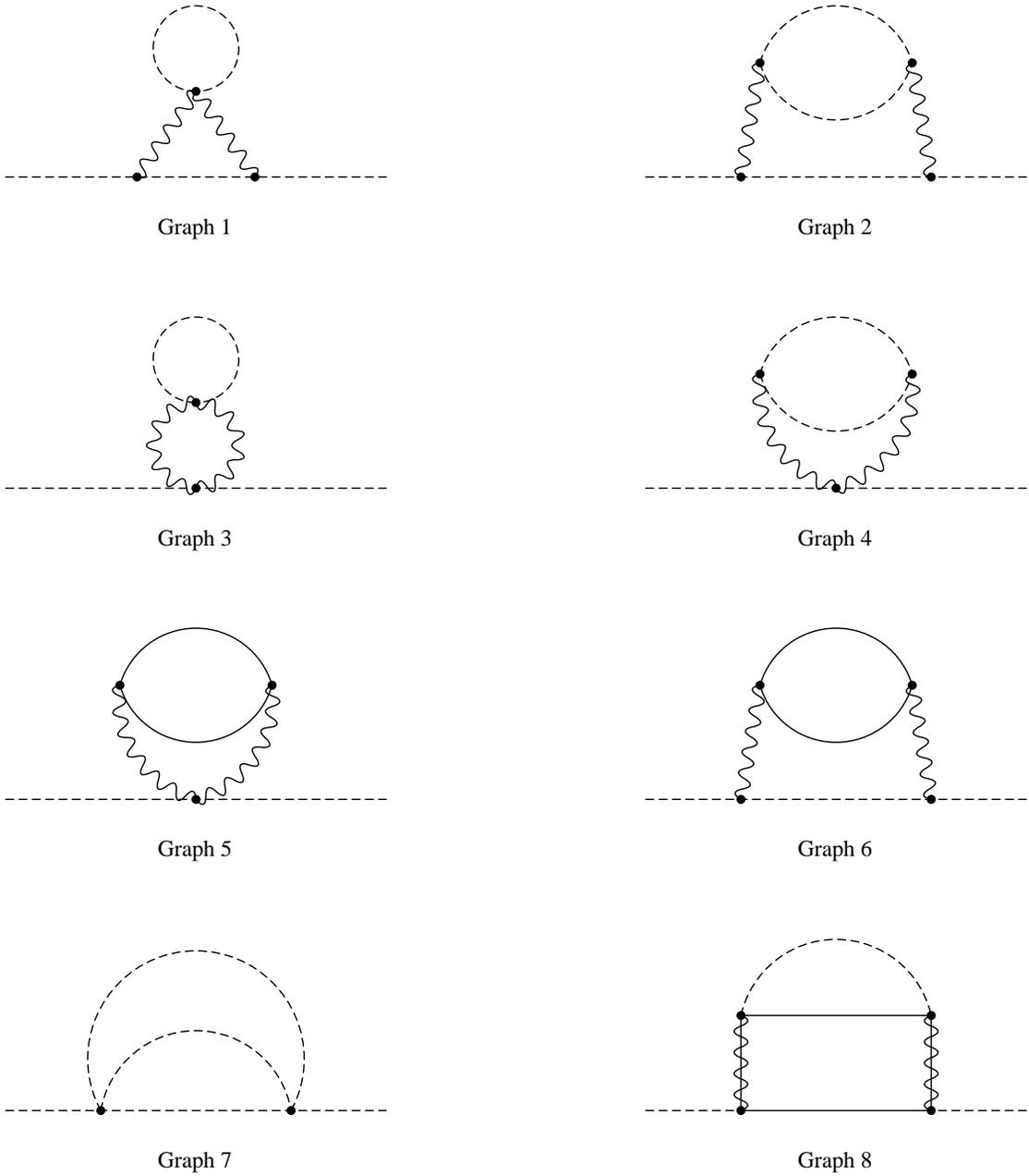}
\caption{Two-loop contributions to MSSM scalar masses
involving messenger sector fields.}
\label{twoloopfigure}
\end{figure}

The scalar masses of the MSSM arise at leading order from 2-loop
graphs shown in Figure \ref{twoloopfigure},
with messenger fields, gauge bosons and
gauginos on the internal lines.
The calculation of these graphs
is described in an Appendix, where we obtain
the result
already given by Dimopoulos, Giudice, and Pomarol\cite{DGP2}:
\beq
{\tilde m}^2=
2 \left |{F/ S} \right |^2 \sum_a \left ({\alpha_a\over 4\pi}
\right )^2 C_a \sum_i n_a(i) f(x_i)
\label{scalarmasses}
\eeq
with
\beq
f(x) =
{1+x\over x^2}\biggl [ \log (1+x) - 2 {\rm Li}_2(x/[1+x])  +
{1\over 2} {\rm Li}_2(2x/[1+x])\biggr ]
+ (x \rightarrow -x) \> .
\label{definef}
\eeq
In (\ref{scalarmasses}), $C_a$ is the quadratic Casimir invariant
of the MSSM scalar field in question, in a normalization where
$C_3=4/3$ for color triplets, $C_2 = 3/4$ for $SU(2)_L$ doublets,
and $C_1 = {3\over 5} Y^2$. It is convenient to write
${\tilde m}^2 = 2 \sum_a ({\alpha_a/ 4\pi})^2 C_a \Lambda_{Sa}^2$
with
\beq
\Lambda_{Sa}^2 =
\left |{F/ S}\right |^2 \sum_i n_a(i) f(x_i) \> .
\eeq
%%%%%%%%%%%%%%%%%%%%%%%%
%%%%%%%%%%%%%%%%%%%%%%%%%%%%%%%%%%%%%%%%%%%%%
In this way the 6 quantities $\Lambda_{Ga}$ and $\Lambda_{Sa}$
parameterize the effects of a non-minimal messenger sector
and non-negligible $x_i$ on the masses of MSSM gauginos and
scalars respectively.
In the limit $|F/\lambda_i S^2| \ll 1$, one recovers the result
$\Lambda_{Ga} = \Lambda_{Sa} =  F/S$ for the minimal model of
\cite{GaugeMediated}, since $f(0)=1$.
In order to illustrate the relative effects of non-negligible
$x_i$ on gaugino and sfermion masses, we graph in Figure \ref{fgplot}
the function $\sqrt{f(x)}$ to compare with $g(x)$.
When $x$ is not very close to 1, one finds excellent precision from
the expansion
\beq
f(x) = 1 + {x^2\over 36} - {11\over 450} x^4 - {319\over 11760} x^6
+\ldots \> .
\eeq
The function $f(x)$ is nearly constant for $x$ not near 1,
and falls sharply near $x=1$ to a minimum value of
$f(1) = 2\log2 + 2\log^2 2 -
\pi^2/6 \approx 0.702$ or $\sqrt{f(1)} \approx 0.838$.
Note that $\sqrt{f(x)}$ is always within one per cent of unity
for $x< 0.85$. Thus as long as $|F/\lambda_i S^2| \lsim 0.85$ for all
messenger fields, one has simply
\beq
\Lambda_{Sa}^2
= |F/S|^2 N_a
\eeq
to a very good approximation.
More generally, one finds
\beq
0.838 {\sqrt{N_a}} \>|F/S| \leq \Lambda_{Sa} \leq  {\sqrt{N_a}} \>|F/S|
\> .
\eeq
By combining the bounds on $g(x)$ and $\sqrt{f(x)}$
we obtain the result
\beq
\sqrt{N_a} \leq {\Lambda_{Ga}\over \Lambda_{Sa}} \leq 1.65 \sqrt{N_a}
\label{ratiobound}
\eeq
in {\it any} model in which all messenger fields obtain their masses only
from a single chiral superfield $S$ and its $F$-term. The effect
of non-negligible $x_i$ is always to lower the masses of squarks
and sleptons compared to the gaugino mass parameters. With some rather
mild restrictions, the range (\ref{ratiobound}) can be
significantly tightened. For example, if all $x_i<0.85$, one can replace
the value $1.65$ by $1.19$. With the further restriction that
all $x_i < 0.5$, the same number becomes only
$1.044$, so that the scales entering the gaugino and scalar
mass formulas differ only at the few per cent level. The 1\%
accuracy level (to which higher-loop corrections are probably
comparable anyway) for $\Lambda_{Ga} \approx {\sqrt{N_a}}
\Lambda_{Sa}$ is reached
if all $x_i < 0.25$.

The masses predicted by equation (\ref{gauginomasses}) and
(\ref{scalarmasses}) are given
at the messenger mass scale(s) 
and must be renormalized down to the scale
of MSSM sparticles. 
Decoupling each set of messengers $\Phi_i, \overline \Phi_i$ at the
appropriate $\lambda_i S$, one obtains
running $\overline{\rm DR}$ gaugino mass parameters
\beq
M_a(Q) = {\alpha_a(Q)\over 4 \pi} {F\over S} \sum_{\lambda_i S > Q}
n_a(i) g(x_i)
\> . \label{decmgaugino}
\eeq
Below the lightest messenger scale this reduces simply to
\beq
M_a(Q) = {\alpha_a(Q) \over 4 \pi} \Lambda_{Ga}
\eeq
up to small two-loop corrections\cite{MV1,Yamada,JJ}.

The scalar (mass)$^2$
parameters obtain renormalization group corrections proportional to
gaugino masses squared, with the result
\beq
{\tilde m}^2(Q) = 2 \sum_a \> C_a \> 
\left (\left | {F\over S} \right |^2 \sum_i
\left [ {\alpha_a(\lambda_i S)\over 4 \pi}\right]^2 
n_a(i) f(x_i) +
\int^{\log \lambda_i S}_{\log Q} d(\log Q^\prime)\> 
{\alpha_a (Q^\prime)\over \pi} M_a^2(Q^\prime)
\right )
\label{sferm}
\eeq
with $M_a(Q)$ given by eq.~(\ref{decmgaugino}), and $\alpha_a(Q)$ by
a similar step-function decoupling of messengers. As long as the couplings
$\lambda_i$ do not feature large hierachies, one may safely neglect
messenger-scale threshold contributions of order $\delta {\tilde m}^2
\sim 2 C_a \log
(\lambda_i/\lambda_j) M_a^2 \alpha_a/\pi$ by choosing a
representative messenger scale $Q_0\approx \lambda_i S$. In this
approximation one finds
\beq
{\tilde m}^2(Q) = 2 \sum_a \left ( {\alpha_a(Q)\over 4 \pi }\right)^2
\>C_a\>
\left [ r_a\Lambda_{Sa}^2 + {1\over b_a}(1-r_a)\Lambda_{Ga}^2
\right ]
\label{renormsfermion}
\eeq
where $(b_1,b_2,b_3) = (-33/5, -1, 3)$ and
\beq
r_a(Q) = [\alpha_a(Q_0)/\alpha_a(Q)]^2 = \left [ 1 + (b_a \alpha_a(Q)/
2 \pi) \log (Q_0/Q) \right ]^{-2} .
\eeq
In the case that all $x_i$ are small and not too different, the
running scalar and gaugino
masses and running gauge couplings can be directly related at any scale by
\beq
{\tilde m}^2 = 2 \sum_a C_a M_a^2
\left (  {r_a\over N_a} + {1\over b_a}(1-r_a)
\right ),
\eeq
while more generally one finds
\beq
{\tilde m}^2 = 2 \sum_a C_a M_a^2
\left (  r_a {\Lambda_{Sa}^2\over \Lambda_{Ga}^2}
+ {1\over b_a}(1-r_a)
\right ),
\label{renormscal}
\eeq
with the ratio ${\Lambda_{Sa}^2 / \Lambda_{Ga}^2}$ bounded
by $0.366/N_a$ and $1/N_a$ according to eq.~(\ref{ratiobound}).
These equations hold at the one-loop level (with Yukawa couplings
and trilinear scalar couplings neglected) in a
non-decoupling $\overline{\rm DR}$ scheme, which means
that MSSM sparticles and Higgs fields are not decoupled at their
mass thresholds.
In order to make precise predictions about the sparticle masses,
these parameters must be related to the physical masses of the particles.
The necessary equations have been given for the gluino and first
and second family squarks in \cite{MV1},
and in general for all of the MSSM particles in \cite{pierce}.

So far we have assumed that the messengers all obtain their masses
entirely through coupling to a single chiral superfield $S$.
If this assumption is relaxed, one clearly obtains a much more
general set of models with a concomitant loss of predictive power.
However, the assumption that only one field $S$ plays a significant
role is perhaps sufficiently compelling that the alternatives can be
considered disfavored. For example, the existence of only one $S$
field succesfully addresses the supersymmetric CP problem, since
all phases in the theory are proportional to the phase of $F/S$,
and can be rotated away. This need not be so if there is more than
one field $S$.
The simplest model of this type is perhaps the obvious extension
of the minimal model of GMSB, i.e.~with messenger
fields $D+\overline D$ and $L+\overline L$, and the superpotential
$
W = \lambda_3 S_3 D\overline D + \lambda_2 S_2 L \overline L
$.
The gaugino masses obtained from this model are given by, in the
small $|F_i/\lambda_i S_i^2|$ limit,
\beq
M_3 = {\alpha_3\over 4\pi} {F_3\over S_3}; \qquad\>\>\>
M_2 = {\alpha_2\over 4\pi} {F_2\over S_2}; \qquad\>\>\>
M_1 = {\alpha_1\over 4\pi}\left ( {2 F_3\over 5 S_3} +
{3 F_2\over 5 S_2} \right ).
\label{interf}
\eeq
If the phases of the VEVs are not aligned, this gives rise to an
observable CP-violating phase arg($M_1 M_2^*$)
which could potentially feed into an electric dipole moment for the
neutron or electron.
On the other hand, if squark and slepton
masses are very large, such
new phases could be tolerable, and the interference in
(\ref{interf}) could allow $M_1$ to be somewhat suppressed relative to
$M_2$ and $M_3$ and the slepton masses.
However, we will not consider such possibilities further here.

\section*{3. Variations in the messenger sector}
\indent

One of the outstanding features of the minimal model of GMSB
is its predictive power,
since the values of the soft supersymmetry-breaking
MSSM parameters are determined by only
a few parameters in the messenger sector.
However one can also entertain the possibility of different field
contents in the messenger sector. The original choice of
messenger fields in ${\bf 5} + {\bf \overline 5}$
of $SU(5)$ is motivated by the fact that it is the
simplest one which simultaneously
provides for plausible
MSSM masses and maintains the apparent unification
of gauge coupling observed at LEP. It is well-known that the
latter feature is shared by any set of
chiral superfields which lie in complete $SU(5)$ GUT multiplets.
The number of such fields which can be used as messenger fields
is then limited by the requirement that the MSSM gauge couplings should
stay perturbative up to the GUT scale
$M_U \approx 2\times 10^{16}$ GeV, which amounts to the
statement
that there can be at most
four ${\bf 5} + {\bf \overline 5}$ sets or one
${\bf 5} + {\bf \overline 5}$ and one ${\bf 10} + {\bf \overline{10}}$.

While maintaining the apparent unification of gauge coupling is
a fine goal, it is not clear how much this really should tell us about
the messenger sector.
First,
it is sufficient but not necessary to have complete
${\bf 5} + {\bf \overline 5}$ and ${\bf 10} + {\bf \overline{10}}$
multiplets of $SU(5)$
in order to maintain perturbative gauge coupling unification.
A counterexample with $N_1=N_2=N_3=3$ is a messenger sector
transforming under $SU(3)_C\times SU(2)_L \times U(1)_Y$ as
\beq
({\bf 3},{\bf 2},{1\over 6}) +
({\bf \overline 3},{\bf 1},{1\over 3}) +
2\times ({\bf 1},{\bf 1},1) + {\rm conj}.
\label{neatomodel}
\eeq
which by itself (or with additional gauge singlets)
does not happen to form any combination of irreducible
representations of any simple GUT group.
Furthermore, it is
not necessary that all TeV or messenger scale
vectorlike chiral superfields
must obtain their masses primarily by coupling to the field $S$.
Those that do not can still
participate in ensuring perturbative gauge coupling unification,
but may not act as messenger fields and in particular can have
little or no effect on the masses of MSSM sparticles.
(There is, after all, a precedent already in the MSSM of chiral
superfields in vectorlike, non-GUT,
representations of the MSSM gauge group without
very large masses, namely the Higgs fields.)
Finally, a skeptic might point out that the apparent unification
of gauge couplings could be partially or wholly accidental,
so that it is prudent to consider equally all alternatives rather
than trust the detailed results of extrapolating coupling
constant relationships over 13 orders of magnitude in energy.

Therefore we will consider here the effects of a somewhat less
constrained messenger sector. We will maintain the constraint that
messenger fields should occupy the same representations as  MSSM
chiral superfields. This is motivated by the fact that stable
messenger particles with exotic $SU(3)_C \times SU(2)_L \times U(1)_Y$
quantum numbers are probably a disaster for cosmology.
In fact it should be noted that in any case the lightest
and the lightest color non-singlet members of the
messenger sector must be stable
insofar as they do not couple to MSSM fields through non-gauge
interactions. (There is an interesting possibility that
a stable neutral messenger might make up the
cold dark matter, however\cite{DGP2}.)
Fortunately, small mixings between non-exotic messengers and their MSSM
counterparts
can allow them to decay; the necessary couplings
may or may not \cite{DNS}
significantly affect the predictions of GMSB. So we consider
five possible types of messenger fields:
\beq
Q+\overline Q = ({\bf 3},{\bf 2},{1\over 6}) + {\rm conj}.; \qquad\qquad
U + \overline U = ({\bf \overline 3},{\bf 1},-{2\over 3}) +{\rm conj}.;\\
D + \overline D = ({\bf \overline 3},{\bf 1},{1\over 3}) +{\rm conj}.;
\qquad\qquad
L+\overline L = ({\bf 1},{\bf 2},-{1\over 2}) + {\rm conj}.;\\
E + \overline E = ({\bf 1},{\bf 1},1) + {\rm conj}.
\eeq
with multiplicities denoted $(n_Q,n_U,n_D,n_L,n_E)$ respectively.
Thus the particle content of the messenger sector
is specified by a five-tuple of integers, given our assumptions.

[Actually, as long as we are only using the numbers  $(n_Q,n_U,n_D,n_L,n_E)$ to
parameterize our ignorance
of non-MSSM physics, we can set
$n_U=0$. This is because the gauge interactions of any
$U+{\overline U}$-type messengers can always be replaced
by messengers in the representations $D+\overline D + E +\overline E$,
as far as the MSSM sector is concerned, since they have the same index for
each group.
Global features of the theory
do depend on $n_U$, of course. One could also consider messenger sectors
which include
single adjoint representations $({\bf 8},{\bf 1},{0})$ or
$({\bf 1},{\bf 3},{0})$, but we will neglect those possibilities here.]

The number of chiral superfields is limited by requiring that gauge couplings
remain perturbative. However, we do not require that the messenger fields
by themselves maintain gauge coupling unification, for the
reasons mentioned above. Instead, we require
as our first criterion only that the messenger fields should be a
subset of some set of
fields that maintains perturbative gauge coupling
unification. Assuming that no messenger field mass greatly exceeds
$10^4$ TeV, the
perturbativity part of the requirement
($\alpha_a \lsim 0.2$ at $M_{\rm Planck}$) amounts to
\beq
N_1 = {1\over 5}(n_Q + 8 n_U + 2 n_D + 3 n_L + 6 n_E) \leq 4
\label{perturb1}\\
N_2 = 3 n_Q + n_L \leq 4
\label{perturb2}\\
N_3 = 2 n_Q + n_U + n_D \leq 4
\label{perturb3}
\eeq
while the full requirement can be written as
\beq
(n_Q,n_U,n_D,n_L,n_E) \leq
(1,0,2,1,2)\>\>\rm{or}\>\>
(1,1,1,1,1)\>\>\rm{or}\>\>
(1,2,0,1,0)\>\>\rm{or}\>\>
(0,0,4,4,0).
\label{unifc}
\eeq
It is possible that the requirements (\ref{perturb1}-\ref{perturb3})
can be weakened, but only
slightly, by allowing the extrapolated gauge couplings to diverge 
between $M_U$ and $M_{\rm Planck}$ or by enlarging the MSSM gauge group
below $M_U$. (Additional
gauge bosons can contribute negatively to the beta functions for
$\alpha_{1,2,3}$, but this effect is limited by constraints on proton
decay, and by the fact that additional chiral superfields which contribute
positively to the beta functions
must also be introduced to break the additional gauge
interactions.)
The requirements that the gluino and the right-handed selectron not
be massless at leading order
imply $N_3 \geq 1$ and $N_1 \geq 1/5$ respectively.
The possibility $N_2=0$ may not be ruled out yet \cite{Faraggi},
if $\tan\beta$ is very small, but it should be decisively
confronted at LEP2
since it requires a chargino mass smaller than $M_W$.
Furthermore, it should be possible to exclude these models with existing
Tevatron data if the decay $\NI \rightarrow \gamma\G$ is prompt, and
perhaps even if it is not.

There are 66
distinct five-tuples $(n_Q,n_U,n_D,n_L,n_E)$
which satisfy these criteria, of which 53 have $N_2\not= 0$.
The number
of distinct combinations $(N_1,N_2,N_3)$ arising from these models
is 40. The ones with $N_2\not=0$ are,
in ascending order of $N_1$:
$
({1\over 5},3,2)
;$ $
({3\over 5},3,3)
;$ $
({4\over 5},4,2)
;$ $
(1,1,1)
;$ $
(1,3,4)
;$ $
({6\over 5},4,3)
;$ $
({7\over 5},1,2)
;$ $
({7\over 5},3,2)
;$ $
({8\over 5},2,1)
;$ $
({8\over 5},4,4)
;$ $
({9\over 5},1,3)
;$ $
({9\over 5},3,3)
;$ $
(2,2,2)
;$ $
(2,4,2)
;$ $
({11\over 5},1,1)
;$ $
({11\over 5},1,4)
;$ $
({11\over 5},3,1)
;$ $
({11\over 5},3,4)
;$ $
({12\over 5},2,3)
;$ $
({12\over 5},4,3)
;$ $
({13\over 5},1,2)
;$ $
({13\over 5},3,2)
;$ $
({14\over 5},2,4)
;$ $
({14\over 5},4,1)
;$ $
({14\over 5},4,4)
;$ $
(3,3,3)
;$ $
({16\over 5},4,2)
;$ $
({17\over 5},1,1)
;$ $
({17\over 5},3,4)
;$ $
({18\over 5},4,3)
;$ $
 ({19\over 5},1,2)
;$ and $
 (4,4,4).
$
The ones with $N_2=0$ and therefore $M_2=0$ at the one loop level are:
$
({2\over 5},0,1)
;$ $
({4\over 5},0,2)
;$ $
({6\over 5},0,3)
;$ $
({8\over 5},0,1)
;$ $
({8\over 5},0,4)
;$ $
({14\over 5},0,1)
;$ $
(2,0,2)
;$ and $
({16\over 5},0,2).
$
%%%%%%%%%%%%%%%%%%%%%%%%%%%%%%%%%%%%%%%%%%%%%%%%%%%%%%%%%%%%%%%%%%%%%
\begin{figure}[tbm]
\centering
\epsfxsize=4.4in
\hspace*{0in}
\epsffile{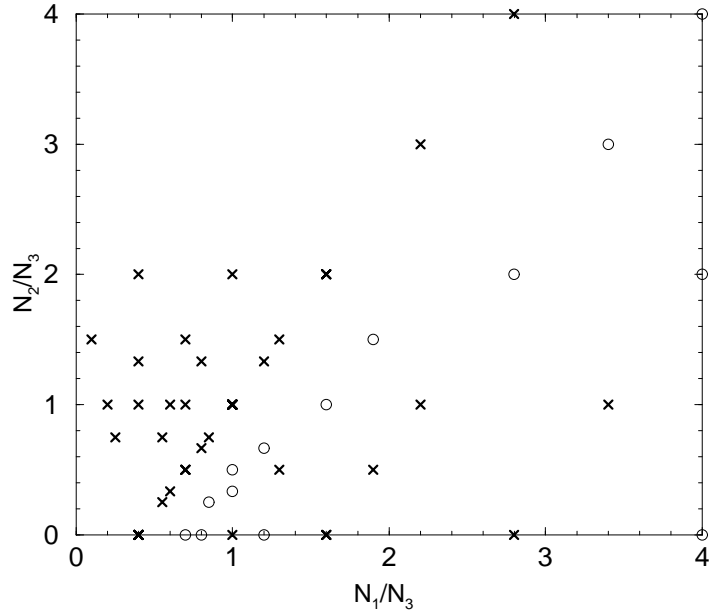}
\caption{A scatterplot of the quantities $N_1/N_3$ and $N_2/N_3$ for
all messenger models satisfying the perturbativity constraints
(\ref{perturb1}-\ref{perturb3}) in the text. Models which do (do not)
also satisfy the ``unification" criteria (\ref{unifc}) are plotted
as Xs (circles).
In the regime
$|F/\lambda_i S^2|\protect\ll 1$, the axis quantities
are equal to the renormalization group scale-independent ratios
$(M_1/\alpha_1)/(M_3/\alpha_3)$ and
$(M_2/\alpha_2)/(M_3/\alpha_3)$ respectively.}
\label{scatterplot}
\end{figure}
%%%%%%%%%%%%%%%%%%%%%%%%%%%%%%%%%%%%%%%%%%%%%

Some indication of the
variety which can be obtained is illustrated in
Figure \ref{scatterplot},
which shows a scatterplot of $N_1/N_3$ vs. $N_2/N_3$
for the 66 models (each shown as an X) which fit the criteria
(\ref{perturb1})-(\ref{unifc}).
For $x_i \ll 1$, these quantities
are equal to the scale-independent gaugino
mass ratios $(M_1/\alpha_1)/(M_3/\alpha_3)$ and
$(M_2/\alpha_2)/(M_3/\alpha_3)$ respectively.
Some of the points on this plot are occupied by several models.
We have also indicated
by circles the presence of 33 models which fit the perturbativity
requirements (\ref{perturb1})-(\ref{perturb3}), but for which
(\ref{unifc}) is not satisfied, so that
the particle content cannot
be embedded into a set which allows perturbative unification of
the gauge couplings unless additional fields with masses far above the
messenger scale are invoked.

The
$n\times({\bf 5} + {\bf \overline 5})$ and
${\bf 10} + {\bf \overline{10}}$
models [and the model in eq.~(\ref{neatomodel})]
all occupy the point
$N_1/N_3 =N_2/N_3 = 1$, but there are other models which
give quite distinctive and interesting predictions.
The models on the $N_2/N_3=0$ axis are the ones with $n_Q=n_L=0$,
which must have small $\tan\beta$ and a chargino lighter than the
$W$ boson; we will omit them from the 
discussions to follow.
The models close to the $N_1/N_3=0$ axis have a
very large hierarchy $m_{\ser} \ll m_{\squark}$, and so may
be strongly disfavored by naturalness criteria.
(We will not attempt here a complete analysis of electroweak
symmetry breaking requirements.) The most ``extreme" such model
has
\beq
(n_Q,n_U,n_D,n_L,n_E) = (1,0,0,0,0);\qquad\qquad (N_1,N_2,N_3) =
({1\over 5},3,2)
\label{extreme}
\eeq
with $M_1$ less than gluino and squark masses by
perhaps a factor of 50, depending on $\alpha_3$.
As can be seen already from Fig.~\ref{scatterplot}, there
can be quite a wide variety in the mass hierarchies between
squarks and gluinos and the sleptons and electroweak gauginos.

\section*{4. Constraints on sparticle masses}
\indent

In this section we will study some features of the sparticle
mass spectrum which follow from the 53 models which satisfy
the constraints
(\ref{perturb1})-(\ref{unifc}) and $N_a>0$ as discussed in the
previous section. We will consider here only the gaugino mass parameters
$M_1$, $M_2$, $M_3$ and the squark and slepton masses of the
first two families, for which Yukawa interactions
can be neglected. We will also not concern ourselves with
the possible origins or role
of the $\mu$ and $B\mu$ terms or scalar trilinear
terms. Constraints following from requiring correct electroweak
symmetry breaking (with viable models for the origins of such terms)
will only further tighten the constraints we will derive.
In the following we assume $F/S < 250$ TeV and $0.01 < x_i < 0.99$.
Taken together these imply that the messenger mass scales are bounded by 
$\lambda_i S < 2.5 \times 10^4$ TeV.

It is perhaps easiest to understand the impact of variations in the
messenger sector by first considering the case that $x_i$ is
small for each messenger pair. In that case the quantities
$\sum_i n_a(i) g(x_i)$ and $\sum_i n_a(i) f(x_i)$ are each
equal to $N_a$, so that the MSSM masses are approximately
determined by just the
parameters $F/S$, $N_1$, $N_2$, $N_3$. Using the values listed above
one can then place some
bounds on the ratios of gaugino mass parameters as follows:
\beq
0.067 {\alpha_1\over\alpha_2} \leq {M_1\over M_2} \leq 3.8
{\alpha_1\over\alpha_2}
\label{ineq1}
\\
0.1 {\alpha_1\over\alpha_3} \leq {M_1\over M_3} \leq 3.4
{\alpha_1\over\alpha_3}
\label{ineq2}
\\
0.25 {\alpha_2\over\alpha_3} \leq {M_2\over M_3} \leq 4
{\alpha_2\over\alpha_3}
\label{ineq3}
\eeq
Although the
gaugino masses run with scale, the veracity of the inequalities
(\ref{ineq1})-(\ref{ineq3}) is renormalization
group scale-independent at one loop. [It is not completely accurate,
however, to replace $\alpha_1, \alpha_2, \alpha_3$ by their measured
values at LEP here, since (\ref{ineq1})-(\ref{ineq3}) hold in the
non-decoupling $\overline{\rm DR}$ scheme.] The lower bounds in
(\ref{ineq1}) and (\ref{ineq2}) are set by the ``extreme" model
in (\ref{extreme}). If this model is discounted, the values $0.067$
and $0.1$ are each raised to $0.2$.

The bounds from (\ref{ineq1})-(\ref{ineq3}) can be strongly modified
by different couplings $\lambda_i$ for messenger fields with different
gauge quantum numbers. However, some general rules can still be found.
For example, we find numerically that $M_1/M_3$ is always
less than 1 at the gluino mass scale, with rough bounds
\beq
0.12 {N_1\over N_3} \lsim {M_1\over M_3} \lsim 0.3{N_1\over N_3}.
\label{m1m3bound}
\eeq
(Note that $N_1/N_3 \leq 3.4$ in these models.)
Also, $M_2$ can only exceed $M_3$ at the gluino mass scale if
$N_2/N_3\geq2$,
and we always find
\beq
0.21 {N_2\over N_3} \lsim {M_2\over M_3} \lsim 0.6{N_2\over N_3}.
\label{m2m3bound}
\eeq
Since $N_2/N_3$ has a maximum value of 4 in these models, the
overall upper limit is $M_2/M_3\lsim 2.4$.
Similarly, $M_1/M_2$ can be as large as about 2.7 at the
electroweak scale, when the $x_i$ are chosen appropriately and $N_1/N_2$
is large. Numerically we find
\beq
0.35 {N_1\over N_2} \lsim {M_1\over M_2} \lsim 0.72{N_1\over N_2}.
\label{m1m2bound}
\eeq

It is interesting to consider the ordering
between the mass of the lightest slepton and the bino mass
parameter $M_1$, since if $|\mu|$ is large, this will give an indication
whether a slepton or a neutralino is the NLSP. 
Using the approximation of eq.~(\ref{renormscal})
one  finds that
\beq
m_{\ser}^2 = {6\over 5} M_1^2 \left [
r_1 {\Lambda_{S1}^2 \over \Lambda_{G1}^2} - {5\over 33}(1-r_1)
\right ]
\label{selapp}
\eeq
for $\overline{\rm DR}$ parameters $m_{\ser}^2$, $M_1$ and $r_1$.
Since ${\Lambda_{S1}^2 \over \Lambda_{G1}^2}\leq 1/N_1$, one finds
that $m_{\ser} > M_1$ can occur only if $N_1 < 66 r_1/(65 - 10 r_1)$.
Now, $r_1$ depends on both the messenger scale and the scale
at which we evaluate the running mass parameters. But a reasonable
estimate for the upper bound is $r_1\lsim 1.7$ [in the regime
of validity of eq.~(\ref{renormscal})], from which
we learn that $m_{\ser} > M_1$ can only
occur if $N_1 \leq 2.2$. This result still applies in more general
situations when eq.~(\ref{sferm}) must be applied. Only 21
models which fit the criteria
of the previous section can satisfy this constraint.
The maximum
values of the ratio $m_{\ser}/M_1$ in these models are approximately
3.0, 1.7, 1.5, and 1.35 for ($n_Q,n_U,n_D,n_L,n_E$) equal to, respectively,
$(1,0,0,0,0)$;
$(1,0,1,0,0)$;
$(1,0,0,1,0)$;
and
$(0,0,1,1,0)$ (the minimal model).
Of course the effect of non-zero $x_i$ can only be to diminish
the ratio $m_{\ser}/M_1$, but the electroweak $D$-term corrections to
$m_{\ser}$ can raise this ratio slightly if $M_1$ is not too large.
There is also a possibility that $M_2$ can be less than both $m_{\ser}$
and $M_1$, if $N_1>N_2$. However, even taking into account the effects
of non-zero $x_i$, we find that this only occurs for a few models with
\beq
(N_1,N_2,N_3) =
({11\over 5},1,1);\>
({11\over 5},1,4);\>
({13\over 5},1,2);\>
({17\over 5},1,1)\>\>
{\rm and}\>\> ({19\over 5},1,2).
\label{largen1models}
\eeq
These are the models for which a line drawn to the origin on Fig.~4
makes the smallest angle with the $N_2/N_3=0$ axis.

If $M_1,M_2 > m_{\ser}$, it is still possible that a neutralino
is the NLSP if $|\mu|$ is not large. This typically means that $\NI$
has a rather large higgsino content, and $\NI\rightarrow\gamma\G$
can be suppressed. However, the competing decays
$\NI\rightarrow h\G$ and $\NI\rightarrow Z\G$ may be kinematically
forbidden, and in any case are subject to very strong kinematic
suppressions $(1-m_{h}^2/m_{\NI}^2)^4$ and $(1-m_{Z}^2/m_{\NI}^2)^4$
respectively\cite{AKKMM2}.
Therefore if $\sqrt{F} < 10^3$ TeV it is still possible
to explain the CDF $ee\gamma\gamma+\Et$ event with small $|\mu|$.
This may be particularly plausible in the chargino interpretation \cite{AKKMM2}
in which the event is due to
$p\overline p \rightarrow \CI\CI$ with allowed two-body decays
$\CI \rightarrow \snu e$
and $\snu \rightarrow \nu\NI$ or $\CI\rightarrow \sel \nu$ and
$\sel \rightarrow e \NI$ followed by $\NI\rightarrow \gamma\G$
in each case. Since the production cross section for chargino
pairs at the Tevatron remains large even for
$m_{\CI} \approx 200$ GeV,
it is sensible to suppose that the two-photon
event could have been seen even
if the decay length of $\NI$ is increased
by a smaller photino component
of $\NI$.

The models in eq.~(\ref{largen1models}) are also interesting because they
minimize the ratio of left-handed to right-handed slepton masses.
In the regime that all $x_i\ll 1$, we find that the running mass parameters
satisfy $m_{\sel}/m_{\ser} \gsim 1.1$ for all of the models
which fit our criteria (with $N_2\geq 1$). The modification of this
ratio due to electroweak $D$-terms happens to be extremely
small because of the
numerical accident $\sin^2\theta_W \approx 1/4$. However, with appropriately
chosen $x_i$, it is possible to obtain $m_{\sel} \approx
m_{\ser}$ in the last two
models of (\ref{largen1models}). 
In all other cases, the hierarchy $m_{\sel} > m_{\ser}$ holds.

One can similarly analyze the possible ranges for the ratios of squark
and gluino masses. It is easiest to consider the particular ratio
$M_{\sdr}/M_{\gluino}$, since this is least sensitive to electroweak
effects. Neglecting the quite small effects of $U(1)_Y$, one finds for
the running mass parameters
\beq
{m_{\sdr}\over M_3}
= {2 \sqrt{2}\over 3} \left [
3 (\Lambda_{S3}/\Lambda_{G3})^2 r_3 + (1-r_3) \right ]^{1/2}
\eeq
in the approximation of eq.~(\ref{renormscal}). 
This ratio is maximized when
$N_3=1$ and all $x_i\approx 0$, and is minimized when
$N_3=4$ and all $x_i\approx 1$. Thus we find
\beq {2 \sqrt{2}\over 3} \left [
1- 0.67 r_3 \right ]^{1/2} \leq
{m_{\sdr}\over M_3}
\leq {2 \sqrt{2}\over 3} \left [
1+2 r_3 \right ]^{1/2}
\> .
\eeq
It is now clear that both the upper and lower limits
are saturated when $r_3$ is as large as possible,
corresponding to a low messenger mass scale. 
Taking estimated bounds
$r_3 \lsim 0.72$ for $N_3=1$ and 
$r_3 \lsim 0.78$ for $N_3=4$, 
we obtain
\beq
0.65 \lsim {m_{\sdr}\over M_3} \lsim 1.48
\qquad\qquad ({\rm running~masses~at}\> Q=m_{\sdr}).
\eeq
Now, the running masses can
be converted into pole masses using the formulas in \cite{MV1,pierce},
yielding a slightly modified estimate for the bounds
on the ratio of the physical pole masses:
\beq
0.66 \lsim {M_{\sdr}\over M_{\gluino}} \lsim 1.36
\qquad\qquad \rm{(pole~masses)}.
\label{sdrglubound}
\eeq
Repeating this type of argument for each value of $N_3$ separately
and taking into account eq.~(\ref{sferm}) one finds
\beq
(0.91,\> 0.76,\> 0.70, \> 0.65 )
\>\lsim \>{M_{\sdr}\over M_{\gluino}} \>\lsim\>
(1.36,\> 1.07,\> 0.95,\> 0.90)
\eeq
for the physical mass ratios with $N_3 = (1,2,3,4)$. The upper limits
in each case correspond to small $x_i$ and small values of
$\lambda_i S$.
The case of left-handed first and second family squarks is slightly
different, especially when $N_2$ is relatively large.
Numerically, we find
\beq
(0.93,\> 0.76,\> 0.70, \> 0.65 )
\>\lsim \>{M_{\sdl}\over M_{\gluino}} \>\lsim\>
(1.74,\> 1.23,\> 1.03,\> 0.95)
\eeq
for the physical mass ratios with $N_3 = (1,2,3,4)$.

The masses of $SU(2)_L$-singlet squarks are never very different from
each other in
the models of section 3. Taking into account the effects of non-zero $x_i$, we
still find a quite narrow range
\beq
1 < m_{\sur}/m_{\sdr} \lsim 1.04
\> .
\eeq
This is not surprising since the $U(1)_Y$ effects are relatively small
even for larger $N_1$. The left-handed squarks are always heavier than
$m_{\sur},m_{\sdr}$. Numerically we find
\beq
1 < m_{\sdl}/m_{\sdr} \lsim (1.1,\> 1.2,\> 1.3,\> 1.4)
\label{sdlsdrbound}
\eeq
for $N_2=(1,2,3,4)$. The squark masses also quite generally exceed
slepton masses even for models with relatively small $N_3$.
Numerically we estimate the bounds
\beq
(1.0,\> 1.5,\> 2.0, \> 2.4 )
\lsim m_{\sdr}/m_{\sel} \lsim
(4,\> 6,\> 8,\> 10)
\label{sdrselbound}
\eeq
for $N_3=(1,2,3,4)$. The situation
$m_{\sdr}\approx m_{\sel}$ only can occur
for $(N_1,N_2,N_3) = ({14\over 5},4,1)$,
the highest point in the plot of Fig.~4.

\section*{5. Discussion}
\indent
In this paper we have examined some of the possibilities for generalized
models of the messenger sector of low-energy supersymmetry breaking.
Despite the large number of discrete model choices and the
freedom to vary the $x_i =|F/\lambda_i S^2|$, the
parameters of the MSSM
are constrained in interesting ways. For example:

\noindent $\bullet$ The usual hierarchy $m_{\ser}\lsim m_{\sel} \lsim
m_{\sdr}\approx m_{\sur} \lsim m_{\sul} \approx m_{\sdl}$
is always preserved, with numerical bounds given
by (\ref{sdrglubound})-(\ref{sdrselbound}).

\noindent $\bullet$ The masses of the right handed squarks
$\sur$, $\sdr$, $\sSr$ and $\sCr$ all lie in a narrow band, and
in a window
within about $\pm35$\% of the physical gluino mass. The upper limit on
$m_{\squark_R}/m_{\gluino}$ is (nearly) saturated for the minimal
${\bf 5}+{\bf \overline 5}$ model with small $x_i$.
The masses of the corresponding left-handed squarks can be significantly
larger in some models, up to about $1.75 m_{\gluino}$.

\noindent $\bullet$ The ratios of
gaugino mass parameters $M_1$, $M_2$, $M_3$
can vary quite
significantly from the predictions of the minimal model, with
$M_2>M_3$ and $M_1>M_2$ both possible at the TeV scale. However,
$M_1/M_3$ is always $\lsim 1$.

\noindent $\bullet$ Only six parameters
$\Lambda_{Ga}$ and $\Lambda_{Sa}$ [plus the overall messenger scale(s)]
enter into the definition of the gaugino mass parameters
and the first and second family squark and slepton masses.
As long as $x_i=|F/\lambda_i S^2|$ is less than about 0.5 (0.25)
for all messenger fields, then there are only four parameters
$F/S$, $N_1$, $N_2$, $N_3$ at the
4\% (1\%) accuracy level, besides a logarithmic dependence on the 
messenger mass scale(s) $\lambda_i S$.

Let us close by noting a slightly different way to express the constraints
on squark and slepton masses which follow from the GMSB framework. One
can see from the form of eq.~(\ref{renormscal}) that 3 parameters suffice
to determine all of the scalar masses for which Yukawa interactions
can be neglected. This means that  for
the 7 scalar masses
$m_{\ser},m_{\snu},m_{\sel},m_{\sdr},m_{\sur},m_{\sul},m_{\sdl}$
there must be 4 sum rules
which do not depend on the input parameters.
Two of these sum rules are completely model-independent
and should hold in {\it any} supersymmetric model (up to small
radiative corrections\cite{Yamada2}):
\beq
m_{\sdl}^2 - m_{\sul}^2 = M_W^2 |\cos 2\beta |; \\
m_{\sel}^2 - m_{\snu}^2 = M_W^2 |\cos 2\beta | .
\eeq
(We assume $\tan\beta>1$.) The other two sum rules can be written
as
\beq
m_{\sur}^2 = m^2_{\sdr} + {1\over 3} m_{\ser}^2 - {4\over 3}
M_Z^2 \sin^2 \theta_W |\cos 2 \beta | ;
\label{sumrule3}
\\
m_{\sdl}^2 = m_{\sdr}^2 + m_{\sel}^2 - {1\over 3} m_{\ser}^2
+ {2\over 3} M_Z^2 \sin^2 \theta_W |\cos 2 \beta |.
\label{sumrule4}
\eeq
These sum rules are not model-independent. It is interesting
to compare with the case of
models with ``supergravity-inspired" boundary conditions featuring
a common $m_0^2$ for scalars and a common $m_{1/2}$ for gauginos
at the GUT or Planck scale. In those models, one finds \cite{MR}
a sum rule which is a particular linear combination
of (\ref{sumrule3}) and (\ref{sumrule4}):
\beq
2 m^2_{\sur} - m^2_{\sdr} - m^2_{\sdl} + m^2_{\sel} - m^2_{\ser}
= -{10\over 3} M_Z^2 \sin^2 \theta_W |\cos 2 \beta |
\label{sumrule5}
\eeq
This sum rule tests the assumption of a common $m_0^2$.
But in GMSB
models, one effectively has the further bit of information
that $m_0^2=0$ (i.e., all contributions to scalar masses are proportional
to the quadratic Casimir invariants; there is no group-independent piece).
This leads to the presence of one additional sum rule, which can be
taken to be either (\ref{sumrule3}) or (\ref{sumrule4}).

It will be an interesting challenge to see to what accuracy these
sum rules can be tested at future colliders.
Perhaps the most interesting possibility
is that the sum rules will turn out to be violated in some gross way; this
would force us to reexamine our assumptions about the origin of
supersymmetry breaking. As an example, suppose that
the messenger sector has some feature which causes additional
unequal supersymmetry-breaking
contributions to the diagonal entries in the mass matrix (\ref{scalmatrix}).
This would lead, through a one-loop graph, to
a Fayet-Iliopoulos $D$-term proportional to weak hypercharge
manifesting itself
in the squark and slepton masses\cite{GaugeMediated}.
Since such a contribution comes in at one loop earlier in the loop
expansion than the contributions from the $F$-term,
it is constrained to be quite small
in order to avoid negative squared masses for some squarks and sleptons.
Conversely, even tiny such
contributions to the matrix (\ref{scalmatrix})
will be magnified in relative importance, and will therefore quite
possibly be observable in the sparticle mass
spectrum! The impact will be to modify  each of the
sum rules (\ref{sumrule3}), (\ref{sumrule4}) and (\ref{sumrule5})
by adding contributions $-4D_Y/3$, $2D_Y/3$ and $-10D_Y/3$ respectively
to the right-hand sides.

In general, we find it remarkable that the models discussed here make such
a  variety of testable predictions. In addition to the possibly dramatic
collider signatures coming from decays of the NLSP into the gravitino,
the sparticle spectrum has a rather distinct character. Future
model building developments will surely tell us even more about what to
expect for the parameters of the MSSM in the low-energy supersymmetry
breaking framework.

\section*{Acknowledgments}
I am indebted to R.~Akhoury, S.~Ambrosanio, T.~Gherghetta,
G.~Kane, G.~Kribs, and S.~Mrenna for helpful discussions.
This work was supported in part by the U.S. Department of Energy.

\section*{Appendix}
In this appendix we give details of the calculation of
the masses of MSSM scalars which arise at leading
order from two-loop diagrams involving messenger fields.
We employ the component
field formalism. There are 8 Feynman diagrams
which contribute at two loop order, as shown in Figure \ref{twoloopfigure}.
We compute these graphs in the Feynman gauge; then one finds that
graph 6 vanishes.
Each of the other graphs is separately divergent but the sum is finite.
It is important to compute all gamma-matrix algebra in 4 dimensions before
computing the momentum integrals with scalar integrands
in $4-2\epsilon$ dimensions, in order
to avoid a spurious non-supersymmetric mismatch between the numbers
of gaugino and gauge boson degrees of freedom.
By straightforward
methods one finds that the contributions for each pair of messenger
fields $\Phi ,\overline\Phi$ are given in terms of the
messenger fermion
mass $m_f=|\lambda_i S|$ and the two messenger scalar masses
$m_{\pm}^2 = |\lambda_i S|^2 \pm |\lambda_i F|$ by:
\beq
\Delta {\tilde m}^2
= \sum_a g_a^4 C_a \>S_a(\Phi)\> \>\>({\rm sum~of~graphs} )
\eeq
where $C_a$ is the quadratic Casimir invariant [normalized
to $(N^2-1)/2N$ for a fundamental representation of $SU(N)$] of the
MSSM scalar,
and $S_a(\Phi)$ is the Dynkin index of
the messenger field $\Phi$ [normalized to
$1/2$ for a fundamental of $SU(N)$].
The contributions to the ``sum of graphs" are given by:
\beq
{\rm Graph}\> 1 =  -{\rm Graph}\> 2 = -{1\over 4} {\rm Graph}\> 3 =
2 \> \langle\mplus\rangle \> \langle 0,0\rangle
+ 2 \> \langle\mminus\rangle \> \langle 0,0\rangle ;
\eeq\beq
{\rm Graph}\>4 =
4 \> \langle\mplus\rangle\>\langle0,0\rangle
+ 4\>  \langle\mminus\rangle\>\langle0,0\rangle
- \>\langle\mplus | \mplus | 0\rangle
- \>\langle\mminus | \mminus | 0 \rangle
\nonumber
\\ - 4 \>\mplus^2 \langle\mplus | \mplus |0,0 \rangle
- 4 \>\mminus^2 \langle\mminus | \mminus |0,0 \rangle
;
\eeq
\beq
{\rm Graph} \> 5 =
8 \> \langle\mf\rangle\>\langle0,0\rangle
- 4 \> \langle\mf | \mf | 0\rangle
+ 8 \> \mf^2 \langle\mf | \mf |0,0 \rangle
;\eeq
\beq
{\rm Graph} \>6 = 0;\qquad\qquad {\rm Graph}\> 7 =
-2\> \langle\mplus | \mminus | 0\rangle ;
\eeq\beq
{\rm Graph} \> 8 =
4 \> \langle\mplus\rangle\>\langle0,0\rangle
+4\>  \langle\mminus\rangle\>\langle0,0\rangle
-8\>  \langle\mf\rangle\>\langle0,0\rangle
+4\>  \langle\mplus | \mf | 0\rangle
+4\> \langle\mminus | \mf | 0  \rangle
\nonumber
\\
+4 (\mplus^2-\mf^2)\> \langle\mplus | \mf | 0 ,0\rangle
+4 (\mminus^2 -\mf^2)\> \langle\mminus | \mf | 0 ,0\rangle .
\eeq
Here we have used the following notation for euclidean momentum integrals
in $n=4-2\epsilon$ dimensions (omitting in each case a factor
$\mu^{2\epsilon}$):
\beq
\langle m \rangle= \int {d^n q\over (2 \pi)^n}
{1\over (q^2 + m^2)}
\eeq
\beq
\langle m,m \rangle = \int {d^n q\over (2 \pi)^n}
{1\over (q^2 + m^2)^2}
\eeq
\beq
\langle m_1 | m_2 |m_3 \rangle =
\int {d^n q\over (2 \pi)^n}
\int {d^n k\over (2 \pi)^n}
{1\over (q^2 + m_1^2)(k^2 + m_2^2)([k-q]^2 + m_3^2)}
\eeq
\beq
\langle m_1 | m_2 |m_3,m_3\rangle =
\int {d^n q\over (2 \pi)^n}
\int {d^n k\over (2 \pi)^n}
{1\over (q^2 + m_1^2)(k^2 + m_2^2)([k-q]^2 + m_3^2)^2}
\eeq
(In the quantities $\langle 0,0\rangle$ and $\langle m_1|m_2|0,0\rangle$, it is
necessary to keep a finite infrared regulator mass
$m_\epsilon$ which will cancel from physical quantities.)

The terms of the form $\langle m \rangle\>\langle 0,0\rangle$ are
easily seen to cancel between the various graphs 1-8, by the
magic of supersymmetry.
The remaining two-loop integrals can be evaluated by standard Feynman
parameter techniques.
First it is convenient to use the identity
\beq
(-1+2 \epsilon)\> \langle m_1 |m_2 |0 \rangle =
m_1^2\>\langle m_2 | 0 |m_1,m_1 \rangle +
m_2^2\>\langle m_1 | 0 |m_2,m_2 \rangle
\eeq
to express everything in terms of dimension 8 integrands. Then
one finds, following e.g.~the methods of \cite{BV}:
\beq
\langle m_1 | m_2 | 0,0 \rangle =
{\Gamma (1+2 \epsilon) \over 2(4 \pi)^n } \Biggl [
{1\over \epsilon^2}
+ {1\over \epsilon}(1 - 2 \log \meps^2) +
1 - \pi^2/6
- F_2(m_1^2,m_2^2) - 2 F_3 (m_1^2, m_2^2)
\nonumber
\\ +  [-2 + 2 F_1(m_1^2, m_2^2)]\log \meps^2 + \log^2\meps^2
\Biggr ] + {\cal O}(\epsilon)
\eeq
and
\beq
\langle m_1 | 0| m_2,m_2 \rangle =
{\Gamma (1+2 \epsilon) \over 2 (4 \pi)^n } \Biggl [
{1\over \epsilon^2}
+ {1\over \epsilon}(1 - 2 \log m_2^2) +
1 - \pi^2/6  - 2 \log m_2^2 + \log^2 m_2^2
\nonumber
\\ - \log^2 m_1^2 +
2 \log m_1^2 \log m_2^2 - 2 {\rm Li}_2(1- m_2^2/m_1^2)
\Biggr ] + {\cal O}(\epsilon)
\eeq
where we have introduced yet more notation:
\beq
F_1(a,b) = (a\log a - b\log b)/(a-b)
\\
F_2(a,b) = (a \log^2 a - b\log^2 b)/(a-b)
\\
F_3(a,b) = [a {\rm Li}_2(1-b/a) - b {\rm Li}_2(1-a/b)]/(a-b)
\eeq
when $a\not=b$, and
\beq
F_1(a,a) = 1 + \log a
\\
F_2(a,a) = 2 \log a + \log^2 a
\\
F_3(a,a) = 2.
\eeq
The dilogarithm or Spence function is defined by
${\rm Li}_2(x) = -\int_0^1 (dt/t) \log(1-xt)$.
Now it is straightforward to add all the contributions to the ``sum
of graphs".
In particular, it is easy to show that the ultraviolet and infrared
divergent terms cancel. The resulting expression can be
simplified further using standard dilogarithm identities \cite{Lewin},
finally yielding the expression given in
\cite{DGP2}, and in equation
(\ref{definef}) of the present paper.

%%%%%%%%%%%%%%%%%%%%%%%%%%%%%%
%  Bibliography
%%%%%%%%%%%%%%%%%%%%%%%%%%%%%%

\end{document}